\documentclass[
  reprint,
superscriptaddress,
aps,
pra,
]{revtex4-2}

\usepackage{graphicx}
\usepackage{dcolumn}
\usepackage{bm}
\usepackage{hyperref}
\usepackage[mathlines]{lineno}
\usepackage{braket}
\usepackage{amsmath}
\usepackage{here}
\usepackage{color}

\begin{document}


\title{Measurement of $p$-wave contacts in an ultracold Fermi gas of $^6$Li atoms}

\author{Kenta Nagase}
\email{Contact author: nagase.k.860b@m.isct.ac.jp}
\author{Soki Oshima}%
\author{Hikaru Takahashi}%
\author{Takashi Mukaiyama}%

 \email{Contact author: mukaiyama@phys.sci.isct.ac.jp}
\affiliation{%
 Department of Physics, Institute of Science Tokyo, Ookayama 2-12-1, Meguro-ku, Tokyo 152-8550, Japan
}%

\date{\today}

\begin{abstract}
Using radio-frequency (RF) spectroscopy, we measure the $p$-wave contacts in a Fermi gas of $^6$Li atoms near the $p$-wave Feshbach resonance. The RF spectrum exhibits clear asymptotic behavior, characterized by $\tilde{\omega}^{-1/2}$ and $\tilde{\omega}^{-3/2}$ dependencies. The magnetic-field dependence of the $p$-wave contacts agrees reasonably well with the virial expansion theory in a detuned magnetic field range, validating the theory near the half the Fermi temperature. The $p$-wave contacts measured in this study constitute a second dataset, complementing the data obtained from the $^{40}$K system and contributing valuable insights into $p$-wave interactions in ultracold Fermi gases.

\end{abstract}

\maketitle


\emph{Introduction}.~Understanding how microscopic few-body interactions characterize the collective behavior of many-body systems is essential for studying strongly correlated fermionic systems. The ability to experimentally control inter-atomic interactions via Feshbach resonances~\cite{FBR} enables an in-depth analysis of the relationship between microscopic interactions and collective properties. Numerous experimental and theoretical studies have unveiled the dynamics of Tan's adiabatic sweep theorem, which links the total energy to the inverse $s$-wave scattering length $a_s$ through a scalar parameter known as the "contact," $C$~\cite{TAN20082952, TAN20082987, TAN20082971}. The impact of the contact is that the value has been demonstrated to relate various thermodynamic quantities with the scattering length. Furthermore, these thermodynamic relations have been proven to hold consistently across diverse conditions, including different particles such as fermions or bosons~\cite{BEC}, different temperature ranges~\cite{homogeneous_contact, temp_dep}, varying interaction strengths~\cite{molecular_fraction, closed_moleculre, Hu_2011}, changing system dimensionalities~\cite{1D,2D_s-wave}, and different states of matter (superfluid or normal). Collectively, these relationships, defined using the contact parameter, are known as universal relations. In addition to these relations, the normalized contact value also shows universality, expressed as a function of $1 / k_{\rm{F}} a_s$~\cite{Hu_2011}, where $k_{\rm{F}}$ denotes the Fermi wavenumber. This implies that determining the contact parameter provides a reliable means to approximate the thermodynamic properties of a correlated system. Previous studies have experimentally verified the universality of the contact parameter across fermionic atoms of $^6$Li and $^{40}$K using various methods~\cite{Bragg, closed_moleculre, molecular_fraction}.

Following the considerable advancements achieved by studies on the universal $s$-wave contact, researchers have now focused their attention on fermions interacting with nonzero angular momentum, specifically through $\ell=1$ ($p$-wave) scattering. Unlike the $s$-wave contact, $p$-wave contacts involve two constants, $C_v$ and $C_R$~\cite{virial, Yoshida, 2D, momentum_p, Inotani, Qi, coexist}. Each of these constants represents the change in total energy associated with variations in the two scattering parameters that characterize $p$-wave interactions: the scattering volume $v$ and the effective range $R$. In 2016, a research group in Toronto reported the first measurement of the two $p$-wave contacts using momentum distribution and radio-frequency (RF) spectroscopy in a $^{40}$K atomic system~\cite{Luciuk2016}. The contact values obtained from these distinct methods demonstrated strong agreement, confirming the existence of $p$-wave contacts. Building upon the progress in understanding the $s$-wave contact, the next step is to investigate the universality of $p$-wave contacts across diverse atomic species. In particular, understanding how the intrinsic differences in the effective range $R$ manifest in the universality of the $p$-wave contact is of significant physical importance. However, except for the measurements conducted by the Toronto group using $^{40}$K atoms, no other experimental studies have yet measured the $p$-wave contacts.

This article reports experimental measurements of the $p$-wave contacts, $C_v$ and $C_R$, in a $^6$Li fermionic system near a 159~G $p$-wave Feshbach resonance~\cite{pFBR1, pFBR2, pFBR3, pFBR4, pFBR_splitting, Nakasuji}. We extract the $p$-wave contacts using conventional RF spectroscopy~\cite{Luciuk2016, verification, homogeneous_contact}, substantially extending the RF pulse length to accurately resolve the small contact values. First, we confirm that the spectral tail at high frequencies follows the expected asymptotic laws, providing clear evidence for species-independent universal relations in $p$-wave contacts. We also investigate the growth dynamics of the $p$-wave contacts, reflecting the local thermalization timescale of the system under strong relaxation. Furthermore, we assess the magnetic-field dependence of the $p$-wave contacts and observe two similarities with previous measurements in the $^{40}$K system: (i) $C_v$ peaks at a magnetic field slightly detuned from the Feshbach resonance, while (ii) $C_R$ peaks at a greater magnetic field detuning than that of $C_v$. Finally, we compare the magnetic-field dependence of the $p$-wave contacts with predictions derived from the virial expansion theory, validating the theory at large magnetic field detuning near the Fermi temperature. The observed $p$-wave contact $C_v$ in the $^6$Li system is one order of magnitude smaller than that in the $^{40}$K system. This finding is consistent with predictions from the virial expansion expression, which suggests that $C_v$ is proportional to the effective range parameter. Our measurements will enable a precise description of the thermodynamic properties of $p$-wave fermions, which have previously been challenging to access due to significant atomic losses.

\emph{Sample preparation}.~Our experiment begins by trapping $^6$Li atoms in a cigar-shaped optical dipole trap (ODT). After 6~s of evaporative cooling in the ODT at 300~G, we obtain a degenerate Fermi gas comprising atoms in the two lowest hyperfine states, $|1\rangle = |F=1/2, m_F=1/2\rangle$ and $|2\rangle = |F=1/2, m_F=-1/2\rangle$, in a 6:4 ratio. This ratio is determined by the initial conditions within the magneto-optical trap. To maximize the number of atoms polarized in the state $|2\rangle$, we flip the spin balance via an RF adiabatic rapid passage and then remove the atoms in state $|1\rangle$. After an adiabatic ramp-up of the trapping depth, we obtain a purified sample containing $N=9(1)\times10^5$ atoms in state $|2\rangle$ in the ODT, with trapping frequencies of $(\omega_x, \omega_y, \omega_z) = 2\pi \times (3874, 3125, 23)$ Hz. The temperature of the gas is approximately $T/T_{\rm{F}} = 0.30$, where $T_{\rm{F}} = 5.6\ \mu\text{K}$ is the Fermi temperature. Some experiments are conducted at a lower Fermi temperature; however, $T/T_{\rm{F}}$ is maintained constant just before RF spectroscopy. To perform RF spectroscopy on the $p$-wave Fermi gas, we adiabatically sweep the magnetic field to the resonance point near the $|1\rangle$-$|1\rangle$ $p$-wave Feshbach resonance at 159 G. Near this resonance, we reduce magnetic field fluctuations to approximately 6 mG through current stabilization.

\begin{figure}[t]
\includegraphics[width=8.5cm]{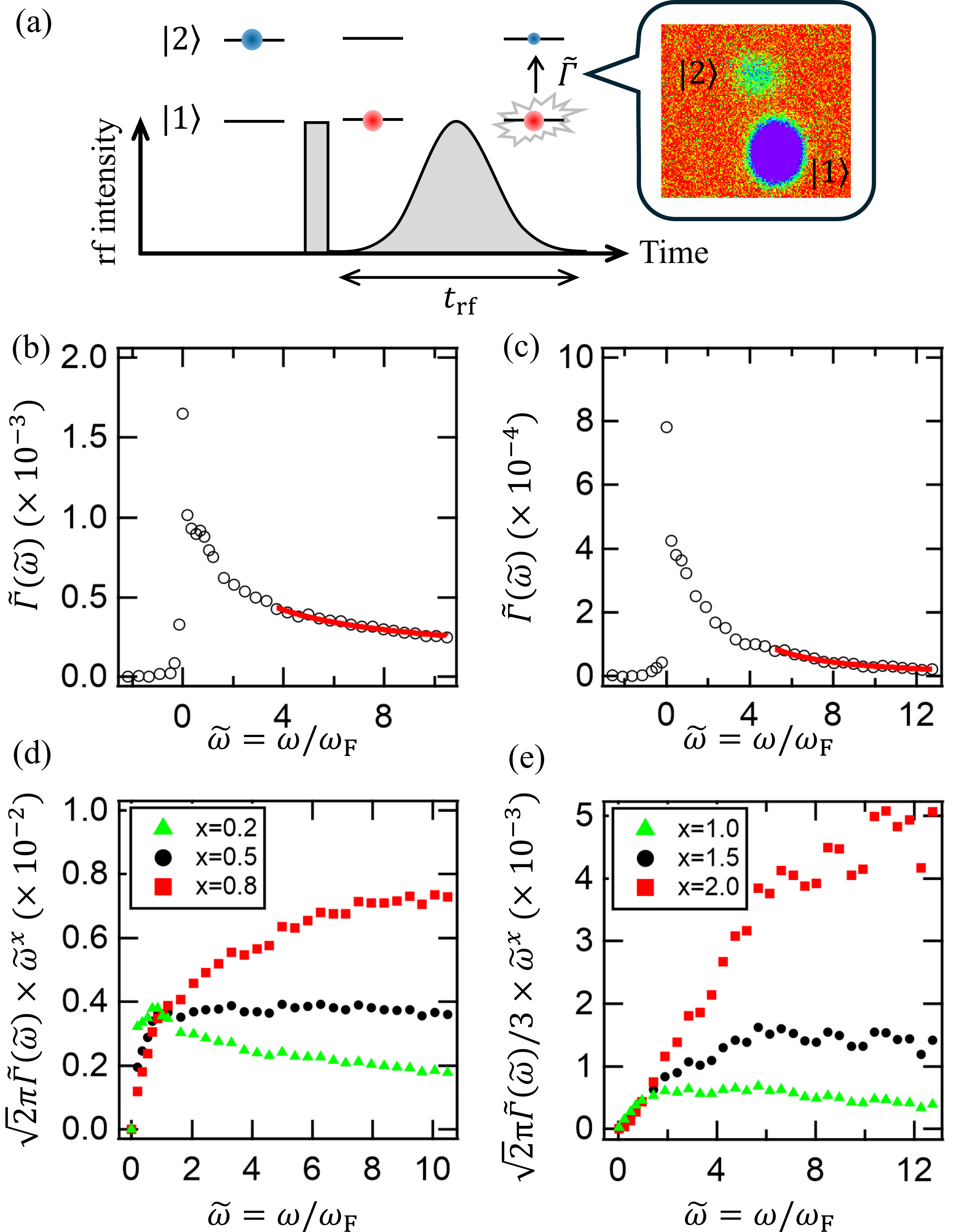}
\caption{\label{RFspec}(a) Experimental time sequence of the two RF pulses. The image displays the result of state-selective absorption imaging after the second pulse. (b, c) RF spectra recorded at $T_{\rm{F}} = 5.6\ \mu\text{K}$ and $\delta B = 41\ \text{mG}$ (b) and at $T_{\rm{F}} = 2.0\ \mu\text{K}$ and $\delta B = 62\ \text{mG}$ (c). Each data point represents an average of five shots. The red lines indicate the best fits obtained using a single power-law function $A \times \tilde{\omega}^x$ in a sufficiently denuded regime, yielding $x = -0.50(2)$ for (b) and $x = -1.56(7)$ for (c). (d, e) $\tilde{\omega}^x$ dependencies of the RF spectra presented in (b) and (c), respectively. The extracted contact values are $(k_{\rm{F}} C_v / N, C_R / k_{\rm{F}} N) = (3.76(9) \times 10^{-3}, 0.00(16) \times 10^{-3})$ for (b) and $(0.00(4) \times 10^{-3}, 1.48(9) \times 10^{-3})$ for (c).
}
\end{figure}

\emph{RF spectroscopy}.~The RF spectroscopy sequence involves two RF pulses, as illustrated in \mbox{Fig.~\ref{RFspec}(a)}. The first RF pulse, with a pulse duration of 5.4 $\mu$s, transfers spin-polarized atoms from state $|2\rangle$ to state $|1\rangle$ ($\pi$-pulse), thereby quenching the $p$-wave interaction. The second RF pulse, applied shortly after the $\pi$-pulse, is longer and has a variable frequency, with a duration of $t_{\rm RF}$. This pulse transfers a fraction of the atoms back to state $|2\rangle$, which is considered free of final-state effects, thus offering insights into the $p$-wave interactions. The second pulse is amplitude-modulated by a Blackman envelope to minimize frequency broadening in the obtained spectrum. Accounting for the reduction in pulse area attributed to the envelope, the maximum attainable Rabi frequency for the second pulse is $\Omega = 2\pi \times 36.5(2)\ \text{kHz}$. Slight variations in the Rabi frequency induced by the characteristics of the RF coil are taken into account when calculating the normalized transfer rate $\tilde{\varGamma}(\tilde{\omega})$~\cite{RabiFrequency}. Following the second pulse, we obtain the transfer fraction $N_2 / N = N_2 / (N_1 + N_2)$ as a function of the second-pulse frequency using state-selective imaging. To avoid spatial overlap of the two spin states in absorption imaging, particularly at high temperatures, we release the atoms from the ODT non-ballistically for 5 ms and subsequently apply a maximum inhomogeneous magnetic field of 110 G/cm for 0.9 ms.

The $p$-wave contact signal appears in the high-frequency tail of the RF spectra. The predicted scaling law for the normalized transfer rate~\cite{Luciuk2016} is expressed as
\begin{equation}
\label{eq1}
\begin{aligned}
\tilde{\varGamma}(\tilde{\omega}) &= \frac{\varGamma(\omega)}{\int_{-\infty}^{\infty} \varGamma(\omega) \, d\omega} \frac{E_{\rm{F}}}{\hbar} \\ 
&\rightarrow  
\frac{1}{\sqrt{2}\pi} \frac{C_v k_{\rm{F}}}{N} \tilde{\omega}^{-\frac{1}{2}} 
+ \frac{3}{\sqrt{2}\pi} \frac{C_R}{k_{\rm{F}} N} \tilde{\omega}^{-\frac{3}{2}}
\end{aligned}
\end{equation}
where $\tilde{\omega} = \hbar \omega / E_{\rm{F}}$, with $E_{\rm{F}}$ denoting the Fermi energy; $\Gamma(\omega) = N_2(\omega) / t_{\rm RF}$; and $\int_{-\infty}^{\infty} \Gamma(\omega) \, d\omega = \pi \Omega^2 N / 2$, based on the sum rule. In this analysis, we assume that the doublet structure of the $p$-wave Feshbach resonance, caused by spin-spin interaction~\cite{splitting_theory, splitting_40K, pFBR_splitting}, is completely overlapped. 
Hence, the $p$-wave contacts represent the sum of the two resonances: $C_v=\sum_{m=0, \pm1}C_{v,m}$ and $C_R=\sum_{m=0, \pm1}C_{R,m}$. Figures~\ref{RFspec}(b, c) display two characteristic RF spectra recorded under conditions where (b) $C_v$ dominates near the resonance and (c) $C_R$ dominates away from the resonance. Notably, these spectra were recorded with a constant RF intensity over a duration of $t_{\rm RF} = 2\ \text{ms}$. The high-frequency tail of each spectrum exhibits the predicted $\tilde{\omega}^{-1/2}$ and $\tilde{\omega}^{-3/2}$ dependencies, as described in Eq.~(\ref{eq1}). In Figs.~\ref{RFspec}(d, e), we plot the measured $\tilde{\Gamma}(\tilde{\omega})$ values, multiplied by $\sqrt{2}\pi \tilde{\omega}^{1/2}$ and $\sqrt{2}\pi/3 \tilde{\omega}^{3/2}$, respectively. The observed plateaus at sufficient detuning, typically $\tilde{\omega} > 5$, provide strong evidence for the universal relations of a Fermi gas with $p$-wave interaction. More importantly, the observation of the same power-law behavior in the RF spectral responses of both the $^{40}$K~\cite{Luciuk2016} and $^6$Li systems suggests that the contact relations are universal across atomic species.

\begin{figure}[t]
\includegraphics[width=8.5cm]{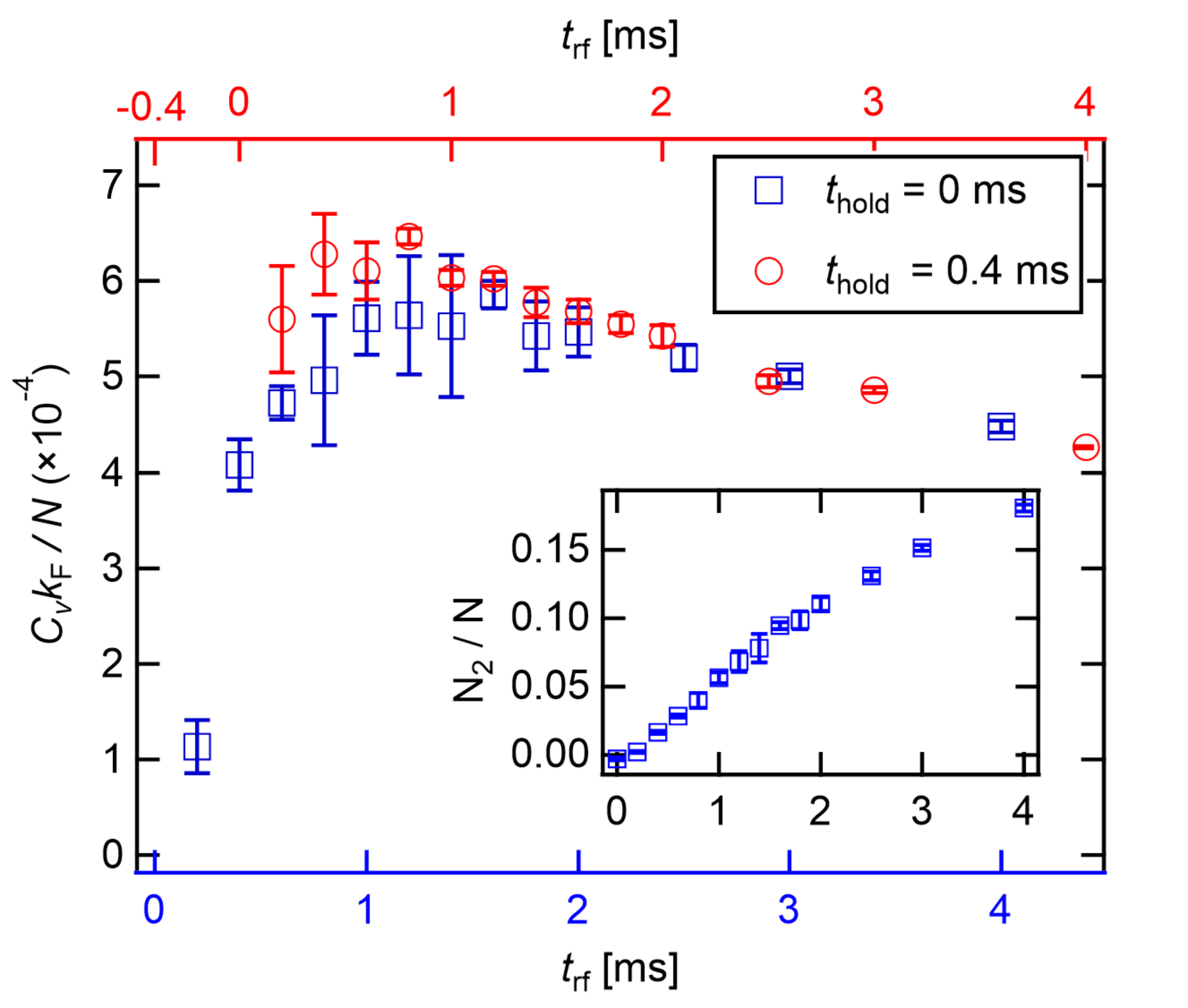}
\caption{\label{hold} Dynamics of the $p$-wave contacts. The value of the normalized contact $C_v k_{\rm{F}} / N$ is determined with varying pulse lengths $t_{\rm RF}$, indicated with no hold time (blue squares) and with a hold time of $t_{\rm hold} = 0.4$ ms (red circles). Each dataset was repeated three times at $T_{\rm{F}} = 2.45\ \mu$K. The magnetic field detuning $\delta B$ was set at 28.8 mG, where the $\tilde{\omega}^{-3/2}$ dependence is absent. Consequently, $C_v$ was extracted from the transfer fraction at a single data point $\tilde{\omega} = 6.44$ in the high-frequency tail. The inset depicts the transfer fraction as a function of $t_{\rm RF}$, indicating a linear increase in the transferred number of atoms up to a pulse duration of 2 ms.}
\end{figure}

\emph{Contact dynamics}.~The contact signal is expected to grow over time as the gas undergoes local thermalization driven by $p$-wave interactions. In Ref.~\cite{Luciuk2016}, the instantaneous growth dynamics of the contact were monitored by varying the hold time $t_{\rm hold}$, which is the interval between the $\pi$-pulse and the spectroscopy pulse. However, under our experimental settings, the spectroscopy pulse is not considered impulsive because the RF pulse duration exceeds the contact growth timescale owing to the RF amplitude limitation. Therefore, we detected the growth dynamics of the $p$-wave contact by varying the RF pulse duration instead of $t_{\rm hold}$.

The blue markers in \mbox{Fig.~\ref{hold}} illustrate the $p$-wave contact $C_v$ as a function of the RF pulse duration at a moderate peak density of $n = 6.0 \times 10^{18}~{\rm m}^{-3}$. We observe that the contact reaches $C_v k_{\text{F}} / N \approx 6 \times 10^{-4}$ at $t_{\rm RF} > 1$~ms. To ensure that this increasing trend of the contact value at short RF pulse durations reflects the growth trend of the contact, we collected the same data with an additional hold time of 0.4~ms, indicated as red markers in Fig.~\ref{hold}. The red data are plotted along the top horizontal axis, shifted 0.4~ms to the right relative to the bottom axis. The consistency between the blue and red data indicates that the RF pulse time and hold time are effectively equivalent. The inset in Fig.~\ref{hold} depicts the number of atoms transferred by the RF pulse as a function of pulse duration, displaying a linear increase up to 2~ms, followed by a slight decay. The decay in contact value at longer pulse lengths occurs because the RF transition dynamics deviates from the linear response regime. Reflecting this result, experiments in Fig.1 and Fig.3 were preformed using RF pulses for less than 2 ms.

\begin{figure}
\includegraphics[width=8.5cm]{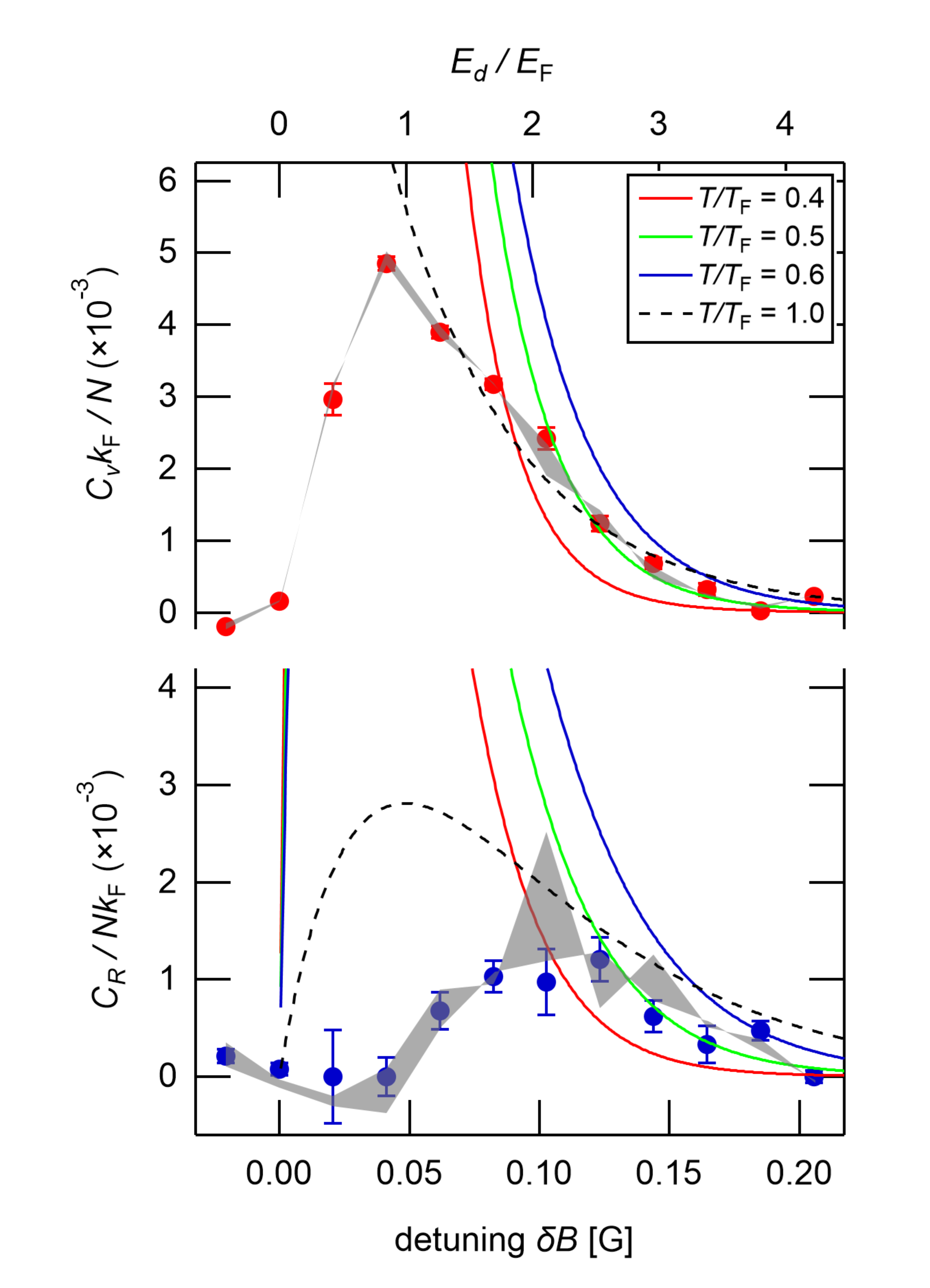}
\caption{\label{Bdep} Magnetic-field dependence of the $p$-wave contacts. The normalized contacts $C_v k_{\rm{F}} / N$ (top) and $C_R / N k_{\rm{F}}$ (bottom) at $T_{\rm{F}} = 5.6~\mu K$ were extracted by fitting in the $\tilde{\omega} > 5$ regime of the spectra. The error bars represent fitting errors, and the shaded area indicates variation in the fitting results based on the selected fitting range, specifically between $\tilde{\omega} > 3$ and $\tilde{\omega} > 7$. Negative values are allowed for the contact, attributed to a systematic offset in the transfer fraction. The solid lines represent theoretical estimates using $n = 2.0 \times 10^{19} \, \text{m}^{-3}$ in the large detuning regime for three distinct temperatures: $T/T_{\rm{F}} =$ 0.4 (red), 0.5 (green), and 0.6 (blue). In contrast, The black dashed line is obtained using measured $n = 3.1\times 10^{18} \, \text{m}^{-3}$ and $T/T_{\rm{F}} = 1.0$  near $E_d/E_{\rm{F}}\approx1$.}
\end{figure}

\emph{Magnetic field dependence}.~Figure~\ref{Bdep} plots $p$-wave contact values measured at various magnetic field detunings from the Feshbach resonance, $\delta B$. Here, we set $t_{\rm RF} = 1.8$~ms to realize an optimal transfer rate for the contact measurement: $t_{\rm RF}$ needs to be longer than the growth time scale but short enough to ensure a linear transfer rate. The change in $\delta B$ is equivalent to tuning the scattering volume $v$, which is parameterized as $v = -v_{\rm bg} \Delta B (1/\delta B)$ near the resonance, with $v_{\rm bg} \Delta B = -2.8(3) \times 10^6 a_0^3$~\cite{Nakasuji}, expressed in terms of the Bohr radius $a_0$. The dimer energy of $p$-wave fermions, normalized by the Fermi energy defined as $E_d / E_{\rm{F}} = 2R / (k_{\rm{F}} v)$, is plotted along the top axis, using an effective range of $R = 11 a_0$~\cite{effective_range}.

We observe that both the contacts $C_v$ and $C_R$ are present only on the Bardeen-Cooper-Schrieffer (BCS) side of the resonance. In particular, the contact $C_v$ sharply increases with detuning, peaking around $\delta B \approx 0.05~\mathrm{G}$. Following this, it monotonically decreases as detuning continues to increase. In contrast, $C_R$ exhibits a small peak at $\delta B \approx 0.1~\mathrm{G}$ before vanishing at larger detuning. Although we observe systematic variations in contact values based on the fitting range (represented by the shaded area), particularly for $C_R$, the overall trend remains consistent. By expressing the data presented on the horizontal axis in a dimensionless form, such as $E_d / E_{\rm F}$, we can directly compare our results for the $^6$Li system with those reported by the Toronto group for the $^{40}$K system~\cite{Luciuk2016}. The magnetic-field dependence of the $p$-wave contacts in both cases is similar: $C_v$ peaks at $E_d / E_{\rm F} \approx 1$, while $C_R$ peaks at $E_d / E_{\rm F} \approx 2$. This suggests the universal feature of $p$-wave contacts near the $p$-wave Feshbach resonance, the origin of which remains an open question for future studies.

In contrast, we observe a difference in the magnetic-field range of the contact signal between the $^{40}$K and $^6$Li systems. Specifically, the measured value of $C_v$~($C_R$) in the $^{40}$K system drops to zero for $E_d / E_{\rm F} > 2~(2.5)$, which can be explained using a closed-channel dimer model. A maximum atom loss of 20\% has minimal effect on the change in $T/T_{\rm F}$, allowing the assumption of well-defined Fermi degeneracy to remain valid. However, in our measurements, the assumption of Fermi degeneracy is not valid because of severe heating of the atoms. In our experiment, the maximum atom loss near the resonance ($\delta B \approx 0.05~\mathrm{G}$) is approximately 70\% when a 1.8 ms RF pulse is applied. In fact, we observe that $T/T_{\rm F}$ rises to approximately one at $E_d / E_{\rm F}\approx1$, at which the assumption of Fermi degeneracy is no longer valid, and reaches $T/T_{\rm F} \approx 0.5$ even in the range of $2 < E_d / E_{\rm F} < 3$. The more significant atomic losses observed for $^{6}\mathrm{Li}$ compared to $^{40}\mathrm{K}$ can be primarily attributed to the smaller contact values in $^{6}\mathrm{Li}$. To sufficiently amplify the contact signals for observation, a longer pulse, which results in greater losses, was required.

Given the deviation from Fermi degeneracy, we explain the behavior of the $p$-wave contacts using a virial expansion approach that is applicable in the high-$T$ regime. The effect of $p$-wave interactions on thermodynamic quantities can be accounted for by the second virial coefficient~\cite{virial,erratum}:
\begin{equation}
\label{b2}
\begin{aligned}
b_2=3 \biggl[ \int_{0}^{\infty}\frac{dk}{\pi}\frac{d\delta(k)}{dk}e^{-\lambda^2k^2/2\pi} + \theta(v)e^{E_b/k_{\rm{B}}T}  \biggr],
\end{aligned}
\end{equation}
where $\delta(k)$ denotes the phase shift, defined as
$k^3 \text{cot}\delta(k)=-1/v-k^2/R$ and $\lambda=h/\sqrt{2\pi m k_{\rm{B}}T}$. We consider the BCS side with $v < 0$, where the two-body bound state is absent. According to the adiabatic theorems, we have
\begin{equation}
\label{Cv}
\begin{aligned}
\frac{C_vk_{\rm{F}}}{N}=24 \sqrt{2} n \lambda k_{\rm{F}} \int_{0}^{\infty}dk\frac{d}{dv^{-1}} \biggl( \frac{d\delta(k)}{dk} \biggr) e^{-\lambda^2k^2/2\pi},
\end{aligned}
\end{equation}
\begin{equation}
\label{CR}
\begin{aligned}
\frac{C_R}{k_{\rm{F}}N}=24 \sqrt{2} \frac{n \lambda}{k_{\rm{F}}}  \int_{0}^{\infty}dk\frac{d}{dR^{-1}} \biggl( \frac{d\delta(k)}{dk} \biggr) e^{-\lambda^2k^2/2\pi}.
\end{aligned}
\end{equation}

Figure~\ref{Bdep} presents the values of the contacts calculated at three temperatures in the $^6$Li system using the following experimentally determined values: $k_{\rm{F}} = 1.2 \times 10^7 \, \text{m}^{-1}$ and $n = 2.0 \times 10^{19} \, \text{m}^{-3}$. Assuming that the virial expansion theory can be applied to the trapped gas down to $T/T_{\rm{F}}\approx0.5$~\cite{Hu_2011}, the observed broadening of $C_v$ and $C_R$ on the higher binding energy side is well explained. As the temperature decreases toward the deeper Fermi-degenerate regime, the width of $C_v~(C_R)$ narrows within $E_d/E_{\rm{F}} = 2~(2.5)$, similar to the behavior observed in the $^{40}$K system. Moreover, both the plots of $C_v$ and $C_R$ shown in \mbox{Fig.~\ref{Bdep}} are consistent with the theory curve with $T/T_{\rm{F}}=0.5$ (green solid curve) at the normalized dimer energy greater than 2. This agrees reasonably to experimentally measured temperature $T/T_{\rm{F}} \approx 0.5$ in the range of $2 < E_d/E_{\rm{F}} < 3$, supporting the validity of the virial expansion theory in a large magnetic field detuning. Near the resonance, however, neither \( C_v \) nor \( C_R \) is explained by the theoretical curve assuming $T/T_{\rm{F}} \approx 0.5$. This may be because of the extensive atom loss and heating of the samples. The black dashed curve shows the theoretical calculation with the peak density of $n=3.1\times10^{18}  \, \text{m}^{-3}$ and the temperature of $T/T_{\rm{F}} = 1.0$, both of which were determined in an independent measurement at $E_d/E_{\rm{F}}\approx1$. While the theoretically obtained $C_v$ value shows agreement with the experimental data near the resonance, the $C_R$ value cannot be explained only by including the atom loss and heating. This seems to indicate a demand for a more precise theory especially in the strongly interacting regime for the $C_R$ contact.

Finally, we discuss the influence of the effective range on the $p$-wave contacts. Notably, both the high-temperature virial expansion and the closed-channel dimer model indicate that $C_v k_{\rm{F}}/N \propto k_{\rm{F}} R$. In the $^6$Li system, the $k_{\rm{F}} R$ value is six times smaller than that of the $^{40}$K atom. In this study, we observed a $C_v$ value that was eight times smaller than that of $^{40}$K, which is consistent with theoretical predictions considering the precision of our measurement. However, as mentioned above, the measurement was done at quite different temperature regimes in our work and the Toronto group. To comprehensively understand the contribution of the effective range to the measured $p$-wave contacts, we must suppress heating effects in our experiment to remain within the Fermi-degenerate regime, enabling a direct comparison with the Toronto group’s results. Given that the RF transition timescale is the limiting factor, we may use another method that minimizes sample heating, such as momentum distribution~\cite{verification, Luciuk2016} or thermodynamic measurements~\cite{verification}.

\emph{Conclusions}.~We measured the $p$-wave contacts in a Fermi gas of $^6$Li atoms using RF spectroscopy. Our results experimentally confirmed that the frequency tails follow two distinct power laws, as predicted by Tan’s relation for $p$-wave contacts. We assessed the magnetic-field dependence of the $p$-wave contacts and compared the obtained results with theoretical predictions from the virial expansion. Our experimental results for the $p$-wave contact $C_v$ agree reasonably well with the theory, supporting the validity of the virial expansion in large magnetic field detuning regions at temperatures near the Fermi temperature. The $p$-wave contact $C_v$ in the $^6$Li system was measured to be one order of magnitude smaller than that in the $^{40}$K system. Notably, the virial expansion predicts that the $p$-wave contact $C_v$ is proportional to the effective range parameter. This is consistent with the fact that the normalized effective range of $^6$Li is six times smaller than that of $^{40}$K, validating the dependence of $C_v$ on the effective range parameter.

Future studies will focus on deriving thermodynamic quantities in systems with strong $p$-wave interactions using the contact relation, potentially deepening our understanding of collective behaviors in these systems. The absolute values of the $p$-wave contacts determined in this work indicate how the energy of the gas changes when the scattering parameters are changed. The concrete and precise physical interpretation of the absolute values will be obtained when the thermodynamic quantities of the gas with strong $p$-wave interactions are measured. Owing to strong $p$-wave inelastic collisions, achieving thermal equilibrium in such systems will be challenging, preventing direct measurements of thermodynamic quantities. However, contact relations provide a promising method for extracting thermodynamic information from locally thermalized states, offering insights into the thermodynamics of systems that only reach local equilibrium. Additionally, measuring $p$-wave contacts and exploring universal contact relations in lower dimensions will be highly valuable for studying non-$s$-wave superfluidity with the impact of dimensionality on system behavior.

\emph{Acknowledgements}.~We acknowledge Y. Chen for the early-stage work and K. G. S. Xie for a discussion on the procedure. The author is supported by the Tsubame Scholarship of the Institute of Science Tokyo. This work was supported by JSPS KAKENHI Grant Number JP24K00553.

\bibliography{apssamp}

\begin{thebibliography}{33}
\bibitem [{Rab()}]{RabiFrequency}%
  \BibitemOpen
  \href@noop {} {}\bibinfo {note} {Our RF coil is designed to have a resonant RF frequency near 68 MHz by directly connecting a resistor and a capacitor to the coil. While this design can maximize the Rabi frequency $\Omega$ at 68 MHz, the Rabi frequency decreases with RF frequency detuning. We find Rabi frequency decreases by approximatedly $2 \pi \times 8$kHz per 1 MHz detuning in RF frequency. To address this, we calibrate the Rabi frequency $\Omega$ across the spectrum range and use the variable $\Omega$ to estimate the transfer rate $\tilde{\Gamma}(\tilde{\omega})$.}\BibitemShut {Stop}%
\end{thebibliography}

\end{document}